\documentclass[preprint,showpacs,preprintnumbers,amsmath,amssymb]{revtex4}

\usepackage{graphicx}
\usepackage{dcolumn}
\usepackage{bm}
\usepackage{amssymb}

\begin{document}
\title{Responses of small quantum systems 
subjected to finite baths
}

\author{Hideo Hasegawa}
\altaffiliation{hideohasegawa@goo.jp}
\affiliation{Department of Physics, Tokyo Gakugei University,  
Koganei, Tokyo 184-8501, Japan}%

\date{\today}

\begin{abstract}
We have studied responses to applied external forces 
of the quantum $(N_S+N_B)$ model for $N_S$-body interacting
harmonic oscillator (HO) system
subjected to $N_B$-body HO bath, by using canonical transformations
combined with Husimi's method for a driven quantum HO 
[K. Husimi, Prog. Theor. Phys. {\bf 9}, 381 (1953)].
It has been shown that the response to a uniform force expressed by the Hamiltonian:
$H_f= -f(t) \sum_{k=1}^{N_S} Q_k$ is generally not proportional to $N_S$
except for no system-bath couplings, where $f(t)$ expresses its time dependence and
$Q_k$ denotes a position operator of $k$th particle of the system.
We have calculated also the response to a space- and time-dependent force expressed by 
$H_f= -f(t) \sum_{k=1}^{N_S} Q_k \: e^{i 2 \pi k u/N_S}$, 
where the wavevector $u$ is $u=0$ and $u=-N_S/2$
for uniform and staggered forces, respectively.
The spatial correlation $\Gamma_m$ for a pair of positions of $Q_k$ and $Q_{k+m}$ 
has been studied as functions of $N_S$ and the temperature.
Our calculations have indicated an importance of taking 
account of finite $N_S$ in studying quantum open systems 
which generally include arbitrary numbers of particles. 

\end{abstract}

        

\maketitle
\newpage
\section{Introduction}
In recent years, there has been considerable interest in open small systems,
whose physical properties have been studied both 
by experimental and theoretical methods \cite{Ritort08}.
We may prepare desired small systems by advanced new techniques.
Theoretical studies of open systems have been made with the use of
the Caldeira-Leggett (CL) type models 
\cite{Ullersma66,Caldeira81,Ford87,Ford88,Weiss99}.
CL-type models have been extensively studied 
by using various methods such as quantum Langevin equation
and master equation \cite{Weiss99}.
The original CL model considers a system of a single
particle ($N_S=1$) which is subjected to a bath consisting of infinite numbers of
uncoupled harmonic oscillators (HOs) ($N_B=\infty$).
Recent studies with the CL model have tried to go beyond this restriction on $N_S$ and $N_B$.
References \cite{Smith08,Wei09,Rosa08} have employed the CL model with 
$N_S=1$ and $N_B \simeq 1 -800$ for studies of properties of small system coupled to finite bath.
CL-type models with $N_S=2$ and $N_B = \infty$ have been investigated \cite{Chou08,Gelin09}.
Reference \cite{Chou08b} discusses the master equation of arbitrary $N_S$ system 
coupled to an arbitrary $N_B$ bath.
In our previous study \cite{Hasegawa11a},
we have adopted the ($N_S+N_B$) model for $N_S$-body system subjected to $N_B$-body bath
in order to calculate energy distributions of a system,
which show intrigue properties as functions of $N_S$, $N_B$ and a system-bath coupling.

In adopting the CL-type model, we have implicitly assumed that 
physical quantities such as the energy and specific heat
of a system with finite $N_S$ ($> 1$)    
are given as $N_S$ times of results of a system with $N_S=1$.
Our recent calculation \cite{Hasegawa11b}, however, has pointed out that
it is generally not the case because 
the system specific heat, $C_S(T;N_S,N_B)$, of the $(N_S+N_B)$ model at temperature $T$ 
is given by
\begin{eqnarray}
C_S(T;N_S,N_B) &\neq & N_S \:C_S(T; 1, N_B),
\end{eqnarray}
except for no system-bath couplings and/or in the high-temperature limit.
Furthermore it has been shown that the low-temperature specific heat may be 
negative for finite $N_S$ with a strong system-bath coupling \cite{Hasegawa11b}. 
This is in contrast with Refs. \cite{Hanggi06,Hanggi08,Ingold09} 
showing a non-negative system specific heat for HO system
in CL-type models with $(N_S,N_B)=(1, 1)$ and $(1, \infty)$.
These results imply that we should explicitly take into account finite $N_S$ 
in studying open systems which may generally include arbitrary numbers of particles. 
It is interesting and necessary to study responses to applied external forces
of the $(N_S+N_B)$ model, 
which is the purpose of the present paper.
Responses of the CL models have been mostly made for infinite baths for which
Ohmic and Drude models are adopted ({\it e.g.}, Ref. \cite{Grabert84}) \cite{Weiss99}.  
In this study, we employ the identical-frequency model for finite baths \cite{Hasegawa11b}.

The paper is organized as follows.
In Sec. II, we briefly explain the $(N_S+N_B)$ model \cite{Hasegawa11a,Hasegawa11b},
to which we apply the canonical transformations 
in order to obtain the diagonalized Hamiltonian including external forces.
By using Husimi's method for a driven quantum HO \cite{Husimi53},
we calculate the response of the open HO system to sinusoidal and step forces.
In Sec. III, we calculate also the response to space- and time-dependent forces.
The spatial correlation $\Gamma_m$ between positions of two particles separated 
by a distance $m$ is evaluated.
The final Sec. IV is devoted to our conclusion.

\section{The ($N_S+N_B$) model}
\subsection{Quantum Langevin equation}
We consider the $(N_S+N_B$) model in which 
the a one-dimensional $N_S$-body system ($H_S$) is subjected to 
an $N_B$-body bath ($H_B$) by the interaction ($H_{I}$) \cite{Hasegawa11a,Hasegawa11b}. 
The total Hamiltonian is assumed to be given by
\begin{eqnarray}
H &=& H_S+H_B+H_{I},
\label{eq:A0}
\end{eqnarray}
with
\begin{eqnarray}
H_S &=& \sum_{k=1}^{N_S} \left[ \frac{P_k^2}{2 M}+ \frac{D Q_k^2}{2}
+ \frac{K}{2}(Q_{k}-Q_{k+1})^2 \right] + H_f, 
\label{eq:A2}\\
H_f &=& -f(t) \: \sum_{k=1}^{N_S} Q_k, 
\label{eq:A2b} \\
H_B &=& \sum_{n=1}^{N_B} \left( \frac{p_n^2}{2 m}
+ \frac{m \omega_n^2 q_n^2}{2}  \right),
\label{eq:A3}\\
H_{I} &=&  \sum_{k=1}^{N_S}  \sum_{n=1}^{N_B} \frac{c_{kn}}{2} (Q_k-q_n)^2.
\label{eq:A4}
\end{eqnarray}
Here $P_k$ ($p_n$) and $Q_k$ ($q_n$) express the momentum and position operators, 
respectively, of a HO with a mass of $M$ ($m$) in the system (bath), 
$D$ and $K$ denote force constants in the system, 
$\omega_n$ is the oscillator frequency of the bath,
$c_{kn}$ is a system-bath coupling and $f(t)$ stands for an applied force.
Operators satisfy commutation relations,
\begin{eqnarray}
[Q_k, P_{\ell}] &=& i \hbar \delta_{k \ell}, \;\;\;
[q_n, p_m] = i \hbar \delta_{n m}, \;\;\;
[Q_k, Q_{\ell}]=[P_k, P_{\ell}]=[q_n, q_m]=[p_n, p_m]=0.
\end{eqnarray}
Equation (\ref{eq:A2}) expresses the interacting HO system 
for $D \neq 0$ and $K \neq 0$.  
In the limiting case of $K=0$, 
the system consists of a collection of uncoupled (independent) HOs.
The system is subjected to a bath 
consisting of a collection of uncoupled HOs with oscillator frequencies of $\{ \omega_n \}$.

In conventional approaches to the quantum system-plus-bath model, 
we obtain equations of motion for $Q_k$ and $q_n$,
employing the Heisenberg equation,
\begin{eqnarray}
i \hbar \dot{O} &=& [O, H],
\end{eqnarray}
where $O$ expresses an arbitrary operator and a dot stands for a derivative
with respect of time.
We obtain the quantum Langevin equations given by \cite{Hasegawa11b}
\begin{eqnarray}
M \ddot{Q}_k(t) &=& -D Q_k(t) -K\left[ 2Q_k(t)-Q_{k-1}(t)-Q_{k+1}(t) \right]
- M \sum_{\ell=1}^{N_S} \xi_{k \ell} Q_{\ell}(t) \nonumber \\
&-& \sum_{\ell=1}^{N_S} \int_0^t \gamma_{k \ell}(t-t') \dot{Q}_{\ell}(t')\:dt'
-\sum_{\ell=1}^{N_S} \gamma_{k \ell}(t) Q_{\ell}(0)+\zeta_k(t)+f(t),
\label{eq:A8} 
\end{eqnarray}
with
\begin{eqnarray}
M \xi_{k\ell} &=& \sum_{n=1}^{N_B}  \left( c_{kn}  \delta_{k \ell} 
-\frac{c_{kn} c_{\ell n}}{m \tilde{\omega}_n^2} \right), 
\label{eq:A9}\\
\gamma_{k\ell}(t) 
&=&\sum_{n=1}^{N_B} \left( \frac{c_{kn} c_{\ell n}}{m \tilde{\omega}_n^2}\right) 
\cos \tilde{\omega}_n t, 
\label{eq:A10}\\
\zeta_k(t) &=& \sum_{n=1}^{N_B} c_{kn} 
\left(q_n(0) \cos \tilde{\omega}_n t
+ \frac{ \dot{q}_n(0)}{\tilde{\omega}_n} \sin \tilde{\omega}_n t \right).
\label{eq:A11}
\end{eqnarray}
Here $\xi_{k \ell}$ denotes the additional interaction between
$k$ and $\ell$th particles in the system induced by couplings $\{ c_{kn} \}$, 
$\gamma_{k\ell}(t)$ stands for the memory kernel and $\zeta_k$ is the stochastic force.
By using averages over initial values of $q_n(0)$ and $\dot{q}_n(0)$,
\begin{eqnarray}
\langle m \tilde{\omega}_n^2 q_n(0)^2\rangle_B
&=& m \langle \dot{q}_n(0)^2 \rangle_B 
=\left( \frac{\hbar \tilde{\omega}_n }{2} \right)
\coth\left( \frac{\beta \hbar \tilde{\omega}_n}{2} \right),
\label{eq:A12}
\end{eqnarray} 
we obtain the fluctuation-dissipation relation,
\begin{eqnarray}
\frac{1}{2}\langle \zeta_k(t)\zeta_{\ell}(t')  +\zeta_{\ell}(t') \zeta_k(t) \rangle_B &=& 
\sum_{n=1}^{N_B} \left( \frac{c_{kn} c_{\ell n}}{m \tilde{\omega}_n^2} \right)
\left( \frac{\hbar \tilde{\omega}_n }{2} \right)
\coth\left( \frac{\beta \hbar \tilde{\omega}_n}{2} \right)
\cos \tilde{\omega}_n (t-t'), \\
&\rightarrow& k_B T \gamma_{k \ell}(t-t')
\hspace{1cm}\mbox{for $\beta \rightarrow 0$},
\label{eq:A13}
\end{eqnarray}
where $\langle \cdot \rangle_B$ expresses the average over initial states of the bath. 
$\xi_{k \ell}$ in Eq. (\ref{eq:A9}) denotes a shift of oscillator
frequency due to an introduced coupling, and it vanishes 
if we adopt $c_{n}=m \tilde{\omega}_n^2$ for $N_S=1$ \cite{Caldeira81}.
In the case of $N_S \neq 1$, however, it is impossible to choose
$\{c_{kn} \}$ such as $\xi_{k \ell}=0$ for all pairs of $(k, \ell)$,
then $Q_k$ is inevitably coupled with $Q_{\ell}$ ($\ell \neq k$).
Because of these couplings between HOs, the $N_S$-body system
cannot be simply regarded as a sum of systems with $N_S=1$.
Although Eqs. (\ref{eq:A8})-(\ref{eq:A11}) are formally exact, 
it is difficult to solve $N_S$-coupled integrodifferential equations.

\subsection{The canonical transformation}
In order to obtain a tractable Langevin equation, we apply the canonical
transformation to the model Hamiltonian.
We assume that $N_S$ is even without a loss of generality. 
Imposing a periodic boundary condition,
\begin{eqnarray}
Q_{N_S+k} &=& Q_{k}, \;\;\; P_{N_S+k}=P_{k},
\label{eq:B1}
\end{eqnarray}
we employ the canonical transformation \cite{Florencio85},
\begin{eqnarray}
Q_k &=& \frac{1}{\sqrt{N_S}} \sum_{s=-N_S/2}^{N_S/2-1}
e^{i (2 \pi k s/N_S)} \:\tilde{Q}_s, 
\label{eq:B2}\\
P_k &=& \frac{1}{\sqrt{N_S}} \sum_{s=-N_S/2}^{N_S/2-1}
e^{i (2 \pi k s/N_S)} \:\tilde{P}_s.
\label{eq:B3} 
\end{eqnarray}
Note that the boundary condition is satisfied in Eqs. (\ref{eq:B2}) and (\ref{eq:B3}) and 
that the set $\{ (1/\sqrt{N_S}) \:e^{i(2 \pi k/N_S) s} \}$ is orthogonal and complete
in a periodic domain of the oscillator label $k$ \cite{Florencio85}.
By the canonical transformation, $H_S$ in Eq. (\ref{eq:A2}) becomes
\begin{eqnarray}
H_S &=& \sum_{s=-N_S/2}^{N_S/2-1} \left[\frac{\tilde{P}_s^* \tilde{P}_s}{2 M}
+ \frac{(D + M \Omega_s^2)  \tilde{Q}_s^* \tilde{Q}_s}{2} \right]
-\sqrt{N_S} \:\tilde{Q}_0 f(t),
\label{eq:B4}
\end{eqnarray}
with
\begin{eqnarray}
M \Omega_s^2 &=& 4K \sin^2 \left(\frac{\pi s}{N_S} \right)
\hspace{1cm}
\mbox{for $s = -\frac{N_S}{2}, -\frac{N_S}{2}+1, \cdot\cdot, \frac{N_S}{2}-1$},
\label{eq:B6}
\end{eqnarray}
where the commutation relations: 
\begin{eqnarray}
[\tilde{Q}_s, \tilde{P}_{s'}^*] = i \hbar \delta_{s s'},\;\;\;
[\tilde{Q}_s, \tilde{Q}_{s'}]=[\tilde{P}_s, \tilde{P}_{s'}]=0,
\end{eqnarray}
hold with $\tilde{Q}_s^*=\tilde{Q}_{-s}$ and $\tilde{P}_s^*=\tilde{P}_{-s}$.

For a simplicity of our calculation, we assume an identical frequency bath 
\cite{Hasegawa11b}, 
\begin{eqnarray}
\omega_n &=& \omega_0, \;\;\;c_{kn}=c.
\label{eq:C1}
\end{eqnarray}
We furthermore assume that $N_B$ is even,
imposing the periodic boundary condition given by
\begin{eqnarray}
q_{N_B+n} &=& q_{n}, \;\;\; p_{N_B+n}=p_{n}.
\label{eq:C2}
\end{eqnarray}
We apply the canonical transformation \cite{Hasegawa11b,Florencio85},
\begin{eqnarray}
q_n &=& \frac{1}{\sqrt{N_B}} \sum_{r=-N_B/2}^{N_B/2-1}
e^{i (2 \pi n r/N_B)} \:\tilde{q}_r, 
\label{eq:C3}\\
p_n &=& \frac{1}{\sqrt{N_B}} \sum_{r=-N_B/2}^{N_B/2-1}
e^{i (2 \pi n r/N_B)} \:\tilde{p}_r,
\label{eq:C4}
\end{eqnarray}
to the bath with the periodic condition given by Eq. (\ref{eq:C2}).
The bath Hamiltonian $H_B$ in Eqs. (\ref{eq:A3}) becomes \cite{Note1}
\begin{eqnarray}
H_B &=& \sum_{r=-N_B/2}^{N_r/2-1} \left(\frac{\tilde{p}_r^* \tilde{p}_r}{2 m}
+ \frac{m \omega_0^2 \tilde{q}_r^* \tilde{q}_r}{2} \right) 
\label{eq:C5} 
\end{eqnarray}
where the commutation relations: 
\begin{eqnarray}
[\tilde{q}_r, \tilde{p}_{r'}^*]=i \hbar \delta_{r r'},\;\;\;
[\tilde{q}_r, \tilde{q}_{r'}]=[\tilde{p}_r, \tilde{p}_{r'}]=0,
\end{eqnarray}
hold with $\tilde{q}_r^*=\tilde{q}_{-r}$ and $\tilde{p}_r^*=\tilde{p}_{-r}$.
By canonical transformations given by Eqs. (\ref{eq:B2}), (\ref{eq:B3}), 
(\ref{eq:C3}) and (\ref{eq:C4}), $H_I$ in Eq. (\ref{eq:A4}) becomes
\begin{eqnarray}
H_I &=& \frac{c N_B}{2} \sum_{s=-N_S/2}^{N_S/2-1} \tilde{Q}_s^*\tilde{Q}_s
+\frac{c N_S}{2} \sum_{r=-N_B/2}^{N_B/2-1} \tilde{q}_r^* \tilde{q}_r
-c \sqrt{N_S N_B}\:\tilde{Q}_0 \:\tilde{q}_0.
\label{eq:C6}
\end{eqnarray}

Summing up Eqs. (\ref{eq:B4}), (\ref{eq:C5}) and (\ref{eq:C6}), 
we obtain the total Hamiltonian expressed by
\begin{eqnarray}
H &=& H_0+ H_S^{'}+H_B^{'},
\label{eq:C7}
\end{eqnarray}
where
\begin{eqnarray}
H_0 &=& \frac{\tilde{P}_0^2}{2M} + \frac{M \tilde{\Omega}_0^2 \tilde{Q}_0^2}{2}
+ \frac{m \tilde{p}_0^2}{2} + \frac{m \tilde{\omega}_0^2}{2}
-c \sqrt{N_S N_B} \tilde{Q}_0 \tilde{q}_0 - \sqrt{N_S} \:\tilde{Q}_0\:f(t),
\label{eq:C8}\\
H_S' &=& \sum_{s (\neq 0)} \left[\frac{\tilde{P}_s^* \tilde{P}_s}{2 M}
+ \frac{M \tilde{\Omega}_s^2  \tilde{Q}_s^* \tilde{Q}_s}{2} \right], 
\label{eq:C9}\\ 
H_B'&=&  \sum_{r (\neq 0)} \left[\frac{\tilde{p}_r^* \tilde{p}_r}{2 m}
+ \frac{m \tilde{\omega}_r^2 \tilde{q}_r^* \tilde{q}_r}{2} \right], 
\label{eq:C10}
\end{eqnarray}
with
\begin{eqnarray}
M \tilde{\Omega}_s^2 &=& D + 4 K \sin^2 \left(\frac{\pi s}{N_S} \right) + c N_B
\hspace{1cm}
\mbox{for $s=-\frac{N_S}{2},\cdot\cdot\cdot, \frac{N_S}{2}-1$}, 
\label{eq:C11}\\
m \tilde{\omega}_r^2 &=& m \omega_0^2
+ c N_S
\hspace{4cm}
\mbox{for $r=-\frac{N_B}{2},\cdot\cdot\cdot, \frac{N_B}{2}-1$}.
\label{eq:C12}
\end{eqnarray}
It is noted that $H_0$ expresses the Hamiltonian
for a uniform mode with $s=u=0$ and that 
a summation over $s$ ($r$) in the $H_S'$ ($H_B'$) 
is excluded for $s=0$ ($r=0$). 

\subsection{Eigenfrequencies with $f(t)=0$}
Eigenfrequencies of the system-plus-bath with $f(t)=0$ 
may be obtained when we diagonalize $H_0$ given by Eq. (\ref{eq:C8}).
We employ the canonical transformation given by
\begin{eqnarray}
\tilde{Q}_0 &=& M^{-1/2}(X_1 \cos \theta+X_2 \sin \theta), \;\;\;\;
\tilde{P}_0 = M^{1/2}(Y_1 \cos \theta+Y_2 \sin \theta), 
\label{eq:C13}\\
\tilde{q}_0 &=& m^{-1/2}(- X_1 \sin \theta+X_2 \cos \theta), \;\;\;
\tilde{p}_0 = m^{1/2}(- Y_1 \sin \theta+Y_2 \cos \theta),
\label{eq:C14}
\end{eqnarray}
where $Y_i=\dot{X}_i$ and their commutation relations are given by
\begin{eqnarray}
[X_i, Y_j] &=& i \hbar \delta_{i j}, \;\;\;
[X_i, X_j]=[Y_i, Y_j]=0
\hspace{1cm}\mbox{for $i, j=1, 2$}.
\end{eqnarray}
The canonical transformation 
yields the diagonalized Hamiltonian given by
\begin{eqnarray}
H &=&H_0 + H_S'+H_B',
\label{eq:C15}
\end{eqnarray}
with
\begin{eqnarray}
H_0 &=&  \frac{Y_1^2}{2}+ \frac{\phi_1^2 X_1^2}{2} 
+ \frac{Y_2^2}{2}+ \frac{\phi_2^2 X_1^2}{2}, \\
\tan 2 \theta &=& \frac{2 c \sqrt{N_S N_B }} 
{ \sqrt{M m} \:(\tilde{\Omega}_0^2 -\tilde{\omega}_0^2)}, 
\label{eq:C16}\\
\phi_{1}^2 &=&
\tilde{\Omega}_0^2 \cos^2 \theta + \tilde{\omega}_0^2 \sin^2 \theta
+\left( \frac{2 c \sqrt{N_S N_B}}{\sqrt{M m} }\right) 
\:\cos \theta \sin \theta, 
\label{eq:C17}\\ 
\phi_{2}^2 &=&
\tilde{\Omega}_0^2 \sin^2 \theta + \tilde{\omega}_0^2 \cos^2 \theta
-\left( \frac{2 c \sqrt{N_S N_B}}{\sqrt{M m} }\right) 
\:\cos \theta \sin \theta,
\label{eq:C18}
\end{eqnarray}
where $H_S'$ and $H_B'$ are given by Eqs. (\ref{eq:C9}) and (\ref{eq:C10}), respectively.
With the use of Eq. (\ref{eq:C16}), $\phi_1^2$ and $\phi_2^2$ are alternatively expressed by
\begin{eqnarray}
\phi_{1, 2}^2 &=& \frac{1}{2}\left[ \tilde{\Omega}_0^2+\tilde{\omega}_0^2
\pm \sqrt{ (\tilde{\Omega}_0^2-\tilde{\omega}_0^2)^2
+ \frac{4 N_S N_B c^2}{M m} } \right],
\label{eq:C19}
\end{eqnarray}
where $+$ ($-$) of a double sign is applied to $\phi_1^2$ ($\phi_2^2$).

In the equilibrium state with $f(t)=0$, Eqs. (\ref{eq:C9}), (\ref{eq:C10}) and (\ref{eq:C19}) 
yield eigenfrequencies of $\{ \nu_i \}$ ($i=1$ to $N_S+N_B$) for $H$ given by
\begin{center}
\begin{tabular}{|c|| c|c|c|c|c|c|c|c|c|c|} \hline
$\;\;i\;\;$ & $1$ &$\cdots$ & $N_S/2+1$ & $\cdots$ & $N_S$ & $N_S+1$ 
& $\cdots$ & $N_S+N_B/2+1$ & $\cdots$ & $N_S+N_B$ \\ \hline
$\;\;\nu_i^2\;\;$ & $\tilde{\Omega}_{-N_S/2}^2$ & $\cdots$ & $\phi_{1}^2$ & $\cdots$
& $\tilde{\Omega}_{N_S/2-1}^2$ & $\tilde{\omega}_0^2$ & $\cdots$ 
& $\phi_{2}^2$ & $\cdots$ & $\tilde{\omega}_0^2$ \\ \hline
\end{tabular}
\end{center}
In the limit of $c=0$, eigenfrequencies become
\begin{center}
\begin{tabular}{|c|| c|c|c|c|c|c|c|c|c|c|} \hline
$\;\;i\;\;$ & $1$ &$\cdots$ & $N_S/2+1$ & $\cdots$ & $N_S$ & $N_S+1$ 
& $\cdots$ & $N_S+N_B/2+1$ & $\cdots$ & $N_S+N_B$ \\ \hline
$\;\;\nu_i^2\;\;$ & $\Omega_{-N_S/2}^2$ & $\cdots$ & $\Omega_0^2$ & $\cdots$
& $\Omega_{N_S/2-1}^2$ & $\omega_0^2$ & $\cdots$ 
& $\omega_0^2$ & $\cdots$ & $\omega_0^2$ \\ \hline
\end{tabular}
\end{center}
Reference \cite{Hasegawa11b} obtained the same eigenfrequencies 
by an alternative method: $\phi_1$ and $\phi_2$ given by Eq. (\ref{eq:C19})
correspond to $\nu_+$ and $\nu_-$, respectively, in Ref. \cite{Hasegawa11b}.
With the use of these eigenfrequencies, the system energy $E_S$
is given by \cite{Hasegawa11b}
\begin{eqnarray}
E_S &=& -\frac{\partial \ln Z_S}{\partial \beta}, \\
&=& \sum_{i=1}^{N_S+N_B} \left(\frac{\hbar \nu_i}{2} \right) 
{\rm coth}\left(\frac{\beta \hbar \nu_i}{2} \right)
-  \left(\frac{N_B \hbar \omega_0}{2} \right) 
{\rm coth}\left(\frac{\beta \hbar \omega_0}{2} \right), 
\end{eqnarray}
where
\begin{eqnarray}
Z_S &=& \frac{Z}{Z_B},
\label{eq:C20}
\end{eqnarray}
with
\begin{eqnarray}
Z &=& {\rm Tr} \;e^{-\beta H}
= \prod_{i=1}^{N_S+N_B} \left[ \frac{1}{2 \sinh(\beta \hbar \nu_i/2 )} \right], 
\label{eq:C21}\\
Z_B &=& {\rm Tr}_B \;e^{-\beta H_B}
= \left[\frac{1}{2 \sinh (\beta \hbar \omega_0/2)} \right]^{N_B},
\label{eq:C22}
\end{eqnarray}
Tr and ${\rm Tr}_B$ denoting a full trace over all variables and
a partial trace over bath variables, respectively.
  
\subsection{Responses to external forces}
\subsubsection{Driven quantum harmonic oscillators}
Quantum HOs driven by an external force have been discussed
in Refs. \cite{Husimi53,Kerner58,Perelomov70}.
It has been shown that the average position of a quantum HO is expressed
by an equation of motion of relevant classical HO \cite{Husimi53,Kerner58,Perelomov70}
as follows.
The Hamiltonian of a single HO 
with mass $m$ and oscillating frequency $\omega_0$ driven 
by a force $F(t)$ is given by \cite{Husimi53,Kerner58,Perelomov70}
\begin{eqnarray}
H &=& \frac{p^2}{2m}+\frac{m \omega_0^2 x^2}{2}-x F(t),
\label{eq:D1}
\end{eqnarray} 
for which the Schr\"{o}dinger equation is expressed by 
\begin{eqnarray}
\left[ -\frac{\hbar^2}{2 m}\frac{\partial^2}{\partial x^2}
+\frac{m \omega_0^2 x^2}{2} -x F(t) \right] \Phi(x,t)
&=& i \hbar \frac{\partial \Phi(x,t)}{\partial t}.
\end{eqnarray}
By using a unitary transformation, we may obtain a solution of $\Phi(x,t)$ expressed by
\cite{Hanggi98}
\begin{eqnarray}
\Phi_n(x,t) &=& \phi_n(x-w(t)) 
\exp\left\{\frac{i}{\hbar} \left[ m \dot{w} (x-w(t))
- E_n t+\int_0^t L(t') \:dt' \right] \right\},
\label{eq:D2}
\end{eqnarray}
with
\begin{eqnarray}
L(t) &=& \frac{1}{2} m \dot{w}^2-\frac{1}{2}m \omega_0^2 w^2+ w F(t), \\
E_n &=& \hbar \omega_0 \left(n+ \frac{1}{2} \right)
\hspace{0.5cm}\mbox{for $n=0,1,2,\cdot,\cdot $}.
\label{eq:D3}
\end{eqnarray}
Here $\phi_n(x)$ and $E_n$ are wavefunction and eigenvalue, respectively,
of the Schr\"{o}dinger equation with $F(t)=0$ in Eq. (\ref{eq:D1}),
and $w(t)$ obeys an equation of motion for a classical driven HO,  
\begin{eqnarray}
m \ddot{w}(t) + m \omega_0^2 w(t) = F(t).
\label{eq:D4}
\end{eqnarray}
Equation (\ref{eq:D2}) shows that the center of a wave packet moves with $w(t)$.
It implies that an average of time-dependent position is given by 
\cite{Husimi53,Kerner58,Perelomov70}
\begin{eqnarray}
\overline{x}(t)=w(t),
\end{eqnarray}
where an overline denotes the quantum average and
$w(t)$ is a solution of Eq. (\ref{eq:D4}).
This is consistent with Ehrenfest's theorem.

\subsubsection{Open quantum system of harmonic oscillators}
In order to study the response of the open quantum HO under consideration,
it is necessary to pursuit equations of classical motions
after Husimi's method \cite{Husimi53,Kerner58,Perelomov70}.
From Eqs. (\ref{eq:C8}) and (\ref{eq:C13}), the total Hamiltonian 
with $f(t) \neq 0$ becomes
\begin{eqnarray}
H &=& H_0+H_S'+H_B',
\label{eq:D6}
\end{eqnarray}
with
\begin{eqnarray}
H_0 &=& H_{01}+H_{02}, 
\label{eq:D7}\\
H_{01} &=& \frac{Y_1^2}{2}+ \frac{\phi_1^2 X_1^2}{2} 
- \sqrt{\frac{N_S}{M}} X_1 \:f(t) \cos \theta, 
\label{eq:D8}\\
H_{02} &=& \frac{Y_2^2}{2}+ \frac{\phi_2^2 X_1^2}{2}
- \sqrt{\frac{N_S}{M}} X_2 \: f(t) \sin \theta,
\label{eq:D9}
\end{eqnarray}
where $H_S'$ and $H_B'$ are given by Eqs. (\ref{eq:C9}) and (\ref{eq:C10}), respectively, 
and $\phi_1$ and $\phi_2$ are given by Eqs. (\ref{eq:C17}) and (\ref{eq:C18}).
Hamiltonians $H_{01}$ and $H_{02}$ in Eqs. (\ref{eq:D8}) and (\ref{eq:D9}) 
express HOs driven by forces of $\sqrt{N_S/M} \:f(t) \cos \theta$ and 
$\sqrt{N_S/M} \:f(t) \sin \theta$, respectively. 
From $H_S'$ in Eq. (\ref{eq:C9}), equations of motion for $\tilde{Q}_s$ with $s \neq 0$ 
are given by
\begin{eqnarray}
M \ddot{\tilde{Q}}_s &=& -M \tilde{\Omega}_s^2 \tilde{Q}_s
\hspace{1cm}\mbox{for $s\neq 0$}
\label{eq:D10}, 
\end{eqnarray}
while Eqs. (\ref{eq:D7}) and (\ref{eq:D8}) lead to
those for $s=0$, $X_1$ and $X_2$, given by
\begin{eqnarray}
\ddot{X}_1 &=& -\phi_1^2 X_1 + \sqrt{ \frac{N_S}{M} } \:f(t) \cos \theta, 
\label{eq:D11}\\
\ddot{X}_2 &=& -\phi_2^2 X_2 + \sqrt{ \frac{N_S}{M} } \:f(t) \sin \theta.
\label{eq:D11b}
\end{eqnarray}
A solution for $\tilde{Q}_0(t)$ may be evaluated from solutions of
$X_1(t)$ and $X_2(t)$ with the canonical transformation given by Eq. (\ref{eq:C13}). 

After some manipulations, quantum-averaged solutions of $\tilde{Q}_s$ are given by
\begin{eqnarray}
\overline{\tilde{Q}}_s(t) &=& \tilde{Q}_s(0) \cos \tilde{\Omega}_s t
+ \frac{\tilde{P}_s(0)}{M \tilde{\Omega}_s} \sin \tilde{\Omega}_s t
\hspace{1cm}\mbox{for $s \neq 0$}, \\
\overline{\tilde{Q}}_0(t) &=&  \tilde{Q}_0(0) A_Q(t) + \tilde{P}_0(0) A_P(t)
+ \tilde{q}_0(0) B_q(t) + \tilde{p}_0(0) B_p(t) + \Phi(t)
\hspace{0.5cm}\mbox{for $s = 0$},
\end{eqnarray}
with
\begin{eqnarray}
A_Q(t) &=& \sum_{i=1}^2 a_i \cos \phi_i t, \\
%
A_P(t) &=& \frac{1}{M} \sum_{i=1}^2 \frac{a_i \sin \phi_i t}{\phi_i}, \\
%
B_q(t) &=& - \sqrt{\frac{m}{M}} \;\cos \theta \sin \theta 
\:(\cos \phi_1 t- \cos \phi_2 t), \\
B_p(t) &=& - \frac{1}{\sqrt{M m}} \:\cos \theta \sin \theta
\left( \frac{\sin \phi_1 t}{\phi_1} -  \frac{\sin \phi_2 t}{\phi_2} \right), \\
\Phi(t) 
&=& \frac{\sqrt{N_S}}{M} \sum_{i=1}^2  \left( \frac{a_i}{\phi_i} \right) \int_0^t
\sin \phi_i(t-t') f(t')\:dt', \\  
a_1 &=& 1-a_2= \cos^2 \theta,
\label{eq:D13}
\end{eqnarray}
where 
$\tilde{Q}_s(0)$, $\tilde{P}_s(0)$, $\tilde{q}_s(0)$ 
and $\tilde{p}_s(0)$ denote initial states.
The response of the total output averaged over initial states is given by
\begin{eqnarray}
R(t) &\equiv& \left< \sum_k \overline{Q}_k(t) \right>_0
= \sqrt{N_S} \left< \overline{\tilde{Q}}_0(t) \right>_0, \\
&=&  \frac{N_S}{M} \sum_{i=1}^2 \left( \frac{a_i}{\phi_i} \right) \int_0^t
\sin \phi_i(t-t') f(t') \:dt',
\label{eq:D12}
\end{eqnarray}
where we employ the relations given by
\begin{eqnarray}
\left< \tilde{Q}_s(0) \right>_0 &=& \left< \tilde{P}_s(0) \right>_0 
= \left< \tilde{q}_s(0) \right>_0 =\left< \tilde{p}_s(0) \right>_0 =0,
\end{eqnarray}
the bracket $\langle \cdot \rangle_0$ expressing an average over initial states. 
Equation (\ref{eq:D12}) leads to the susceptibility,
\begin{eqnarray}
\chi(t) &=& \frac{N_S}{M} \sum_{i=1}^2 \frac{a_i \sin \phi_i t}{\phi_i},
\end{eqnarray}
whose Fourier transformation is given by
\begin{eqnarray}
\hat{\chi}(\omega) &=& \frac{N_S}{M} \sum_{i=1}^2 \frac{a_i}{(\phi_i^2-\omega^2)},
\end{eqnarray}
with poles at $\omega=\pm \phi_i$.

It should be noted that $R(t)$ in Eq. (\ref{eq:D12}) is generally not proportional to $N_S$ 
except for the $c=0$ case 
because $\phi_i$ and $a_i$ depend on $N_S$ 
as shown in  Eqs. (\ref{eq:C17}), (\ref{eq:C18}) and (\ref{eq:D13}). 
This point will be shortly demonstrated in numerical model 
calculations for sinusoidal and step forces in the following.

\begin{figure}
\begin{center}
\includegraphics[keepaspectratio=true,width=100mm]{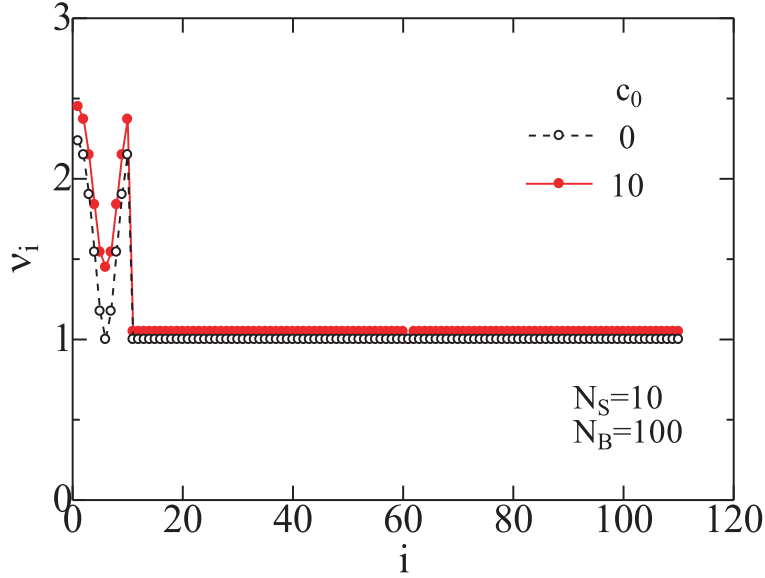}
\end{center}
\caption{
(Color online) 
Eigenfrequencies $\nu_i$ of HO systems with $N_S=10$ 
subjected to a bath with $N_B=100$ for $c_0=0.0$ (open circles)
and $c_0=10.0$ (filled circles) ($D=K=M=m=1.0$ and $\omega_0=1.0$), 
solid and dashed curves being plotted only for a guide of the eye.
}
\label{fig1}
\end{figure}

\begin{figure}
\begin{center}
\includegraphics[keepaspectratio=true,width=120mm]{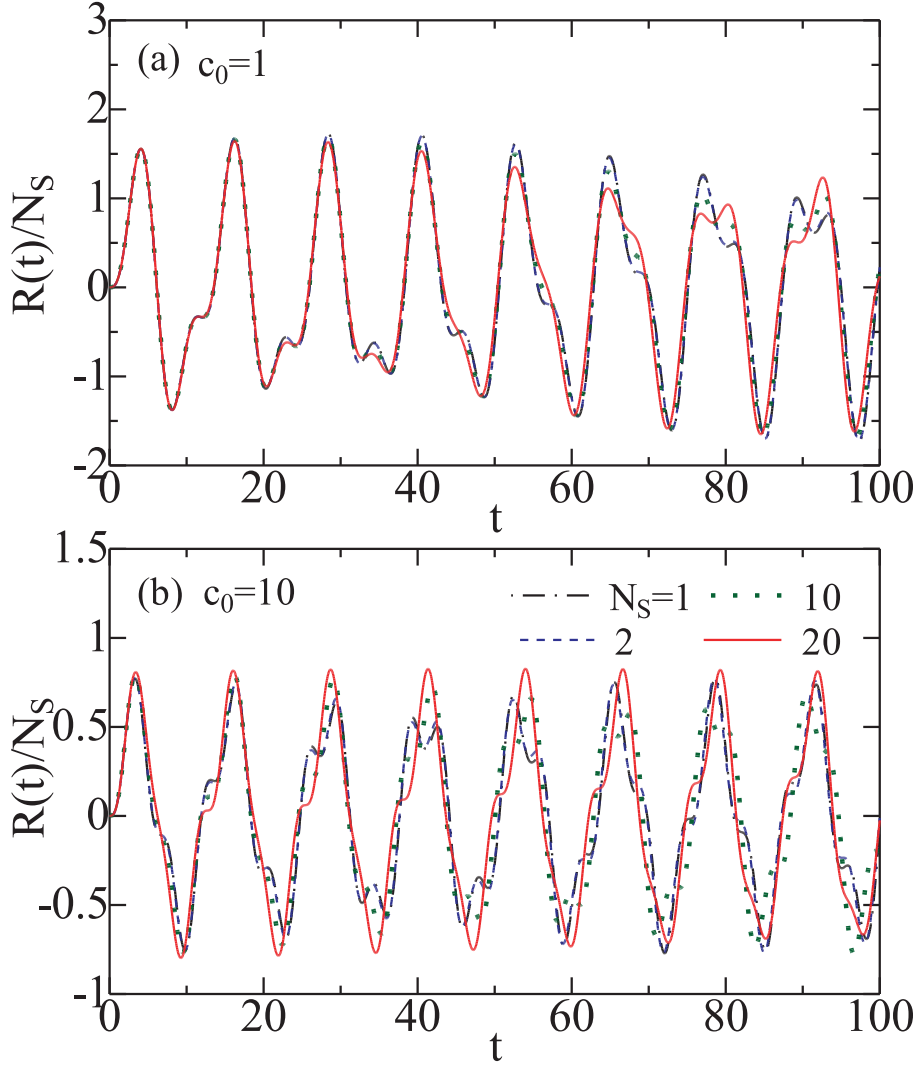}
\end{center}
\caption{
(Color online) 
Responses of $R(t)/N_S$ of HO systems with $N_S=1$ (chain curves), 
2 (dashed curves), 10 (dotted curves) and 20 (solid curves)
for (a) $c_0=1.0$ and (b) $c_0=10.0$
to an applied sinusoidal force with $\omega=0.5$ and $g=1.0$
($D=K=M=m=1.0$, $\omega_0=1.0$ and $N_B=100$).
}
\label{fig2}
\end{figure}

\begin{figure}
\begin{center}
\includegraphics[keepaspectratio=true,width=120mm]{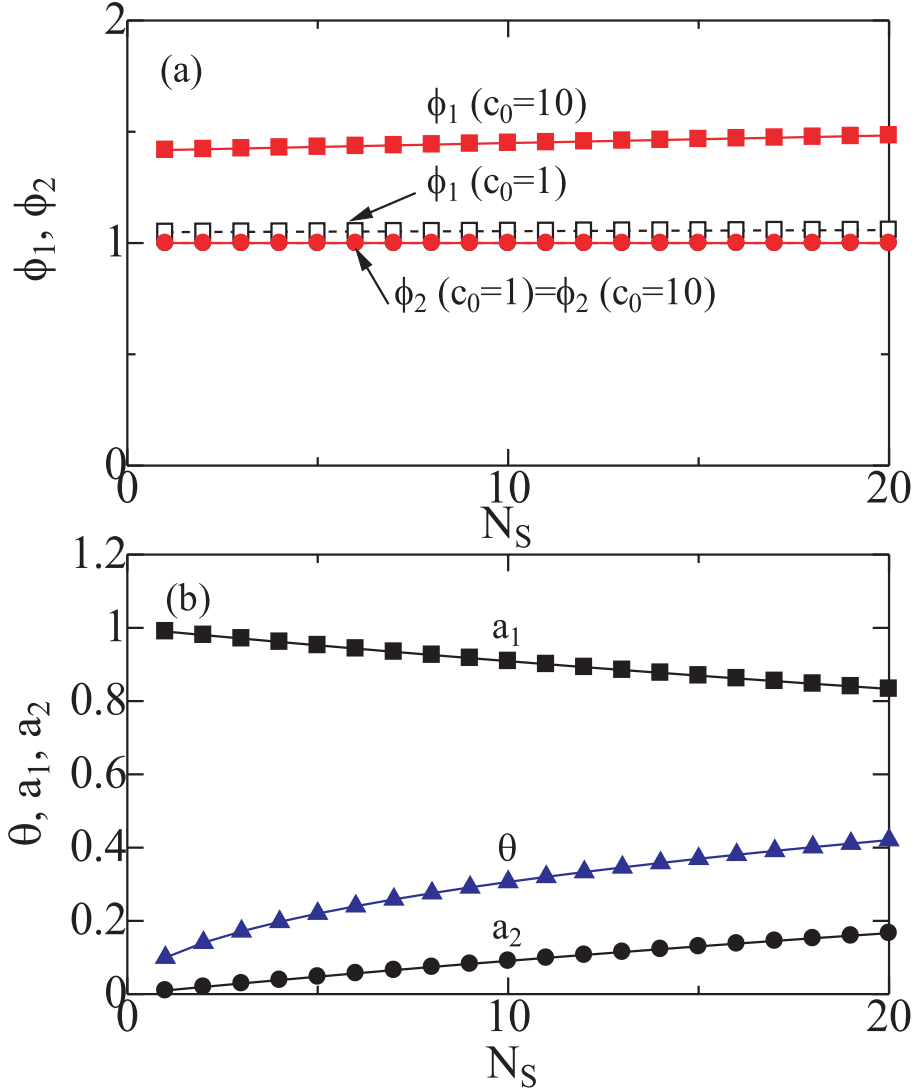}
\end{center}
\caption{
(Color online) 
$N_S$ dependences of (a) $\phi_i$, and (b) $\theta$ and $a_i$ ($i=1,2$)
for $c_0=1.0$ (dashed curves) and 10.0 (solid curves) 
($D=K=M=m=\omega_0=1.0$ and $N_B=100$).
$\theta$ and $a_i$ in (b) are independent of $c_0$ 
for the adopted parameters (see the text).
}
\label{fig3}
\end{figure}

\vspace{0.5cm}
\noindent
{\it A. Sinusoidal forces}

We apply a periodic monochromatic force,
\begin{eqnarray}
f(t) &=& g \sin \omega t,
\label{eq:E1}
\end{eqnarray}
where $\omega$ and $g$ stand for the frequency and magnitude, respectively, of the force.
Equations (\ref{eq:D12}) and (\ref{eq:E1}) yield
\begin{eqnarray}
R(t) &=& \left( \frac{N_S \:g}{M} \right) \sum_{i=1}^2 
\frac{a_i \: (\phi_i \sin \omega t-\omega \sin \phi_i t)}{\phi_i(\phi_i^2-\omega^2)}
\hspace{1cm}\mbox{for $\omega \neq \phi_i$}.
\label{eq:E2}
\end{eqnarray}
In the resonant case of $\omega=\phi_1$, $R(t)$ is given by
\begin{eqnarray}
R(t) &=& \left( \frac{N_S \:g}{M} \right)
\left[ \frac{a_1 \:(\sin \omega t-\omega t \cos \omega t)}{2 \omega^2}
+ \frac{a_2 \: (\phi_2 \sin \omega t-\omega \sin \phi_2 t)}{\phi_2(\phi_2^2-\omega^2)}
\right].
\label{eq:E3}
\end{eqnarray}
Expressions of $R(t)$ in the resonance cases of $\omega = \phi_2$ and $\omega=\phi_1=\phi_2$ 
are similarly given.
In the limit of $c=0$ where $\phi_1=\Omega_0$, $\phi_2=\omega_0$,
$\theta=0.0$, $a_1=1.0$ and $a_2=0.0$, Eq. (\ref{eq:E2}) reduces to
\begin{eqnarray}
R(t) &=& \left( \frac{N_S \:g }{M} \right) 
\left[ \frac{\Omega_0 \sin \omega t- \omega \sin \Omega_0 t}
{\Omega_s(\Omega_0^2-\omega^2)} \right]
\hspace{1cm}\mbox{for $c = 0$ and $\omega \neq \Omega_0$},
\end{eqnarray}
which expresses the response of a HO isolated from a bath.

We have performed numerical model calculations, choosing a coupling
\cite{Hasegawa11b},
\begin{eqnarray}
c &=& \frac{c_0}{N_S N_B},
\end{eqnarray}
such that the interaction term in Eq. (\ref{eq:A4}) including summations over
$\sum_{k=1}^{N_S}$ and $\sum_{n=1}^{N_B}$ yield finite contributions
even in the limits of $N_S \rightarrow \infty$ and/or $N_B \rightarrow \infty$.
We have adopted parameters of $D=K=M=m=\omega_0=1.0$ for a given system-plus-bath.

Figure \ref{fig1} shows eigenfrequencies $\nu_i$ for $c_0=0.0$ (open circles)
and $c_0=10.0$ (filled circles) of a HO system ($N_S=10$) subjected 
to a bath ($N_B=100$). 
Eigenfrequencies $\nu_i$ for $1 \leq i \leq 10$ show a dispersion relation of the HO system 
while those for $11 \leq \nu_i \leq 110$ of the bath are almost constant.
For $c_0=0.0$, we obtain $\tilde{\Omega}_0=1.0$ and $\tilde{\omega}_0=1.0$.
When the system-bath coupling of $c_0=10.0$ is introduced,
they become 1.414 and 1.048, respectively, which lead to
$\phi_1=1.449$ and $\phi_2=1.0$.

Figure \ref{fig2}(a) shows responses of $R(t)/N_S$ to a sinusoidal force 
with $\omega=0.5$ and $g=1.0$
of HO systems with $N_S=1$, 2, 10 and 20 coupled to 
$N_B=100$ baths with a coupling of $c_0=1.0$.
Results of $R(t)/N_S$ are almost the same independently of $N_S$,
although some discrepancies among the four results are realized at $t \gtrsim 50$.
These discrepancies become more evident for a larger coupling of $c_0=10.0$,
whose results are shown in Fig. \ref{fig2}(b).
These results clearly suggest
\begin{eqnarray}
R(t;N_S) &=& N_S R(t;1)
\hspace{1cm}\mbox{for $c=0$}, \\
&\neq& N_S R(t;1)
\hspace{1cm}\mbox{for $c \neq 0$}.
\end{eqnarray}  

In order to elucidate $N_S$ and $c_0$ dependences of $R(t)/N_S$, 
we show in Figs. \ref{fig3}, $\phi_i$, $\theta$ and $a_i$ ($i=1,2$) 
as a function of $N_S$ for $c_0=1.0$ (dashed curves) and 10.0 (solid curves).
Figure \ref{fig3}(a) shows that with increasing $N_S$, $\phi_1$
is slightly increased while $\phi_2$ is constant.
We note in Fig. \ref{fig3}(b) that an increase of $N_S$ yields an increase in
$\theta$, by which $a_2$ is increased but $a_1$ is decreased.
For adopted parameters, $\theta$, $a_1$ and $a_2$ are independent of $c$
because the denominator of Eq. (\ref{eq:C16}) becomes
$\tilde{\Omega}_0^2-\tilde{\omega}_0^2=(N_B-N_S)c$ 
whose $c$ is cancelled out by that in its numerator. 
With increasing $N_S$, a contribution from a lower eigenfrequency of $\phi_2$ 
is increased.
The effect of the system-bath coupling for $c_0=10.0$ is more significant 
than that for $c_0=1.0$
because the difference of $\phi_1-\phi_2$ in the former is larger than 
that in the latter: if $\phi_1=\phi_2$ results are independent of
$a_i$ (and then $N_S$).

\vspace{0.5cm}
\noindent
{\it B. Step forces}

\begin{figure}
\begin{center}
\includegraphics[keepaspectratio=true,width=120mm]{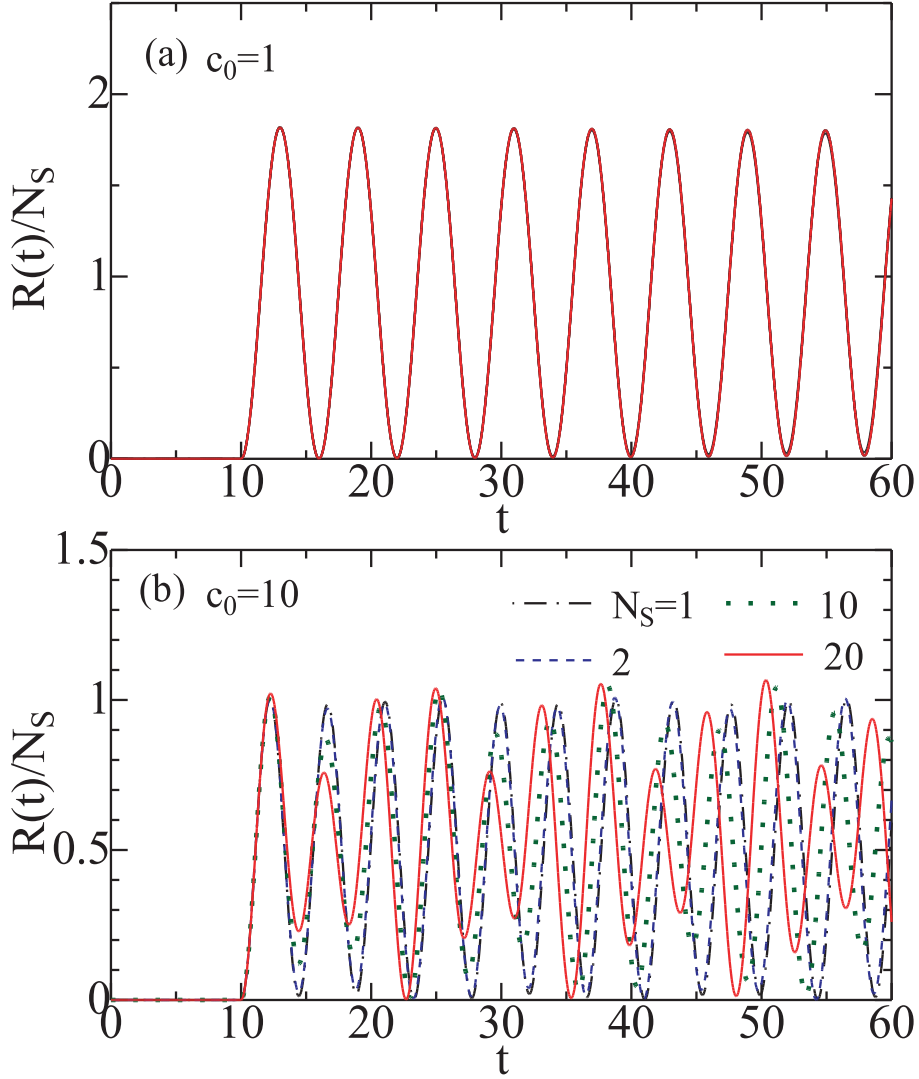}
\end{center}
\caption{
(Color online) 
Responses of $R(t)/N_S$ of HO systems with $N_S=1$ (chain curves), 2 (dashed curves), 
10 (dotted curves) and 20 (solid curves) for (a) $c_0=1.0$ and (b) $c_0=10.0$
to an applied step force with $t_s=10.0$ and $g=1.0$
($B=K=M=m=1.0$, $\omega_0=1.0$ and $N_B=100$). 
Results for all $N_S$ in (a) are indistinguishable.
}
\label{fig4}
\end{figure}

Next we apply a step force given by
\begin{eqnarray}
f(t) &=& g \:\Theta(t_s-t),
\label{eq:E4}
\end{eqnarray}
where $\Theta(x)$ stands for the Heaviside function and
$t_s$ is the starting time of a force with a magnitude of $g$.
The averaged output is given by 
\begin{eqnarray}
R(t) &=& \left( \frac{N_S \:g }{M} \right) \sum_{i=1}^2
\frac{a_i [1-\cos \phi_i(t-t_s)]}{\phi_i^2}.
\end{eqnarray}

Figure \ref{fig4}(a) shows $R(t)/N_S$ for a step force with $t_s=10.0$ and $g=1.0$
of HO systems with $N_S=1$, 2, 10 and 20 coupled to 
$N_B=100$ baths with a coupling of $c_0=1.0$ ($D=1.0$ and $\omega_0=1.0$).
Result of $R(t)/N_S$ for $N_S \geq 2$ are almost the same as that for $N_S=1$.
However, when the interaction is increased to $c_0=10.0$,
the discrepancy between results of $N_S=1$ and $N_S \geq 2$ become evident.
Fig. \ref{fig4}(b) shows similar plots but with stronger coupling of $c_0=10.0$,
for which shape and magnitude of $R(t)/N_S$ are significantly modified
for $N_S \geq 2$.

\section{Discussion}
\subsection{Responses to space- and time-dependent forces}

\begin{figure}
\begin{center}
\includegraphics[keepaspectratio=true,width=120mm]{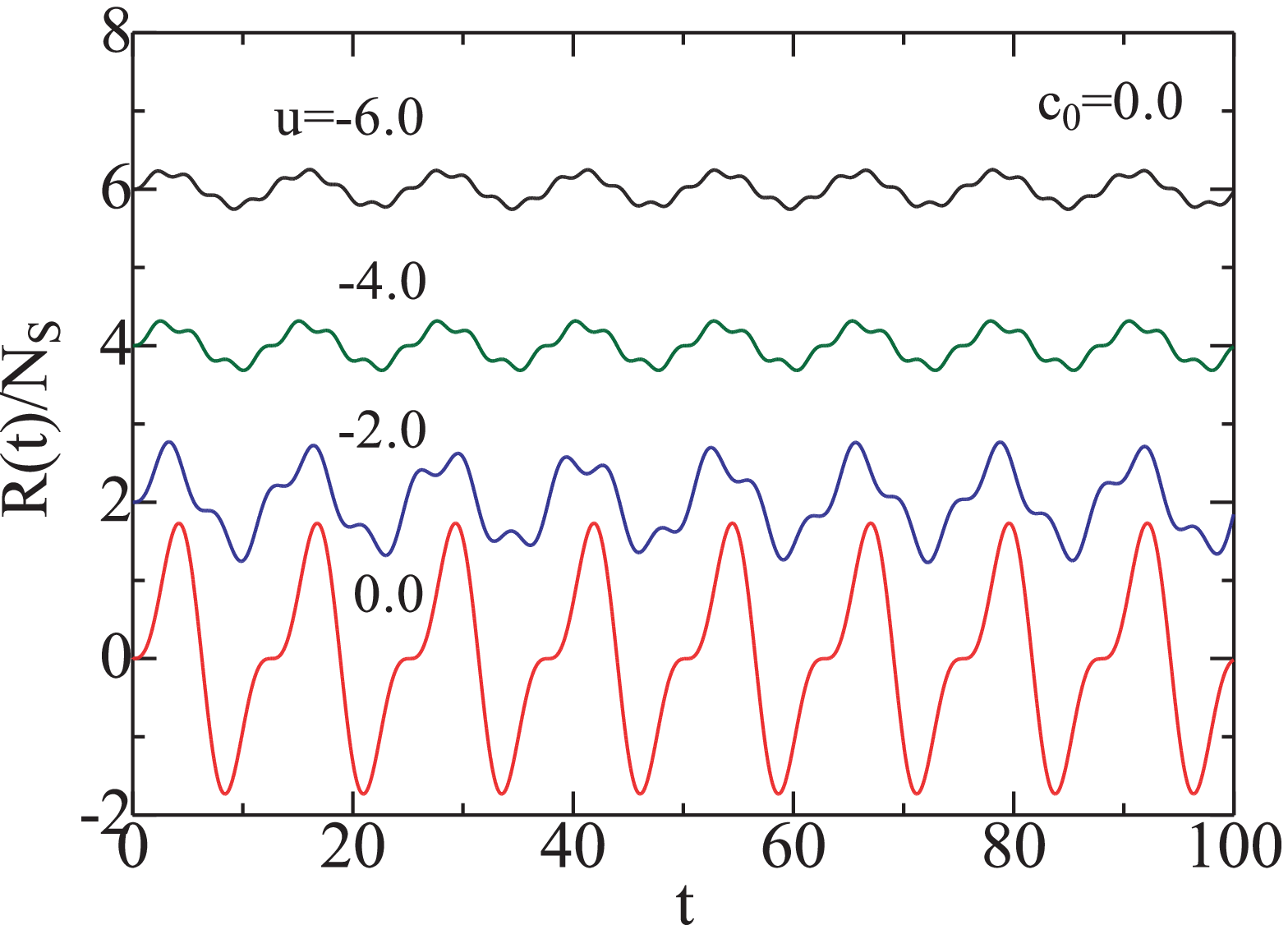}
\end{center}
\caption{
(Color online) 
Responses of $R(t)/N_S$ of an isolated HO system ($c_0=0.0$) 
to an applied sinusoidal force with $\omega=0.5$ and $g=1.0$
for various $u$ ($N_S=12$, $N_B=100$ and $K=D=M=m=\omega_0=1.0$).
Results for $u=-6.0$, $-4.0$ and $-2.0$ are shifted by 6.0, 4.0 and 2.0, respectively,
for clarity of the figures.  
}
\label{fig5}
\end{figure}

\begin{figure}
\begin{center}
\includegraphics[keepaspectratio=true,width=120mm]{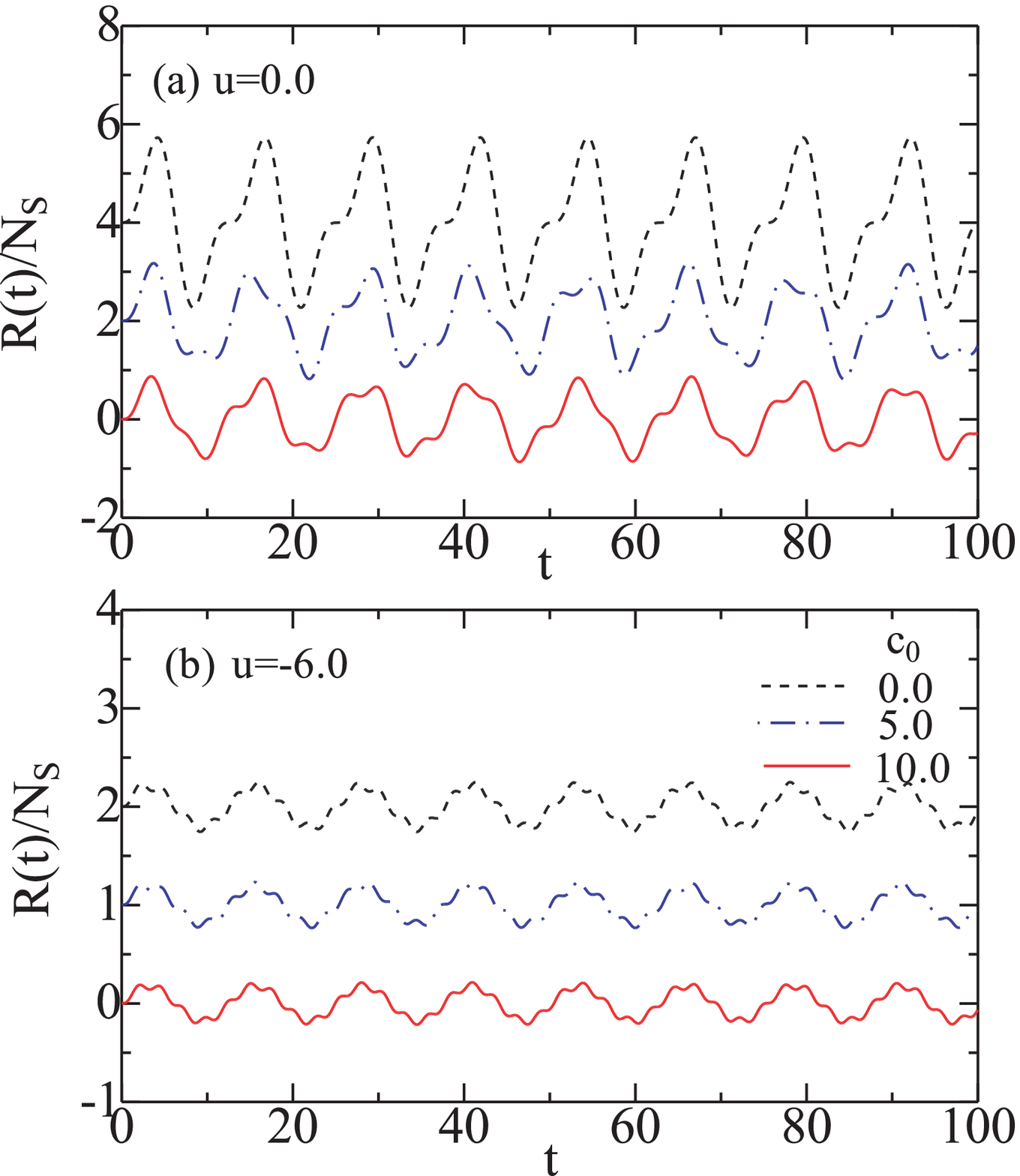}
\end{center}
\caption{
(Color online) 
Responses of $R(t)/N_S$ of HO systems coupled with $c_0=0.0$ (dashed curves), 
$c_0=5.0$ (chain curves) and $c_0=10.0$ (solid curves)
to an applied sinusoidal force with $\omega=0.5$ and $g=1.0$ for
(a) $u=0.0$ and (b) $u=-6.0$ 
($N_S=12$, $N_B=100$ and $K=B=M=m=\omega_0=1.0$).
Results for $c_0=0.0$, $5.0$ in (a) are shifted by 4.0 and 2.0, respectively,
and those for $c_0=0.0$, $5.0$ in (b) are similarly shifted by 2.0 and 1.0,
for clarity of the figures.
}
\label{fig6}
\end{figure}

It is interesting to calculate responses to a space- and time-dependent force
which yields $H_f$ in Eq. (\ref{eq:A2}),
\begin{eqnarray}
H_f &=& - f(t) S(u),
\end{eqnarray}
with 
\begin{eqnarray}
S(u) &=& \sum_{k=1}^{N_B} Q_k \;e^{i 2 \pi k u/N_S}
\hspace{0.5cm}
\mbox{for $u \in \{-\frac{N_S}{2}, -\frac{N_S}{2}+1, \cdot\cdot, \frac{N_S}{2}-1$} \}.
\end{eqnarray}
Here the wavevector $u$ is, for example, $u=0$ and $u=-N_S/2$ for
uniform and staggered forces, respectively, for which $S(u)$ is represented by
\begin{eqnarray}
S(u) &=& \sum_{k=1}^{N_S} Q_k 
\hspace{2cm}\mbox{for $u=0$}, \\
&=& \sum_{k=1}^{N_S} Q_k \;e^{-i \pi k}
\hspace{1cm}\mbox{for $u=-\frac{N_S}{2} $}.
\end{eqnarray}

The mode with $u \neq 0$ does not couple with $s=0$ mode which couples
with bath as mentioned in the preceding subsection II D. 
Equations of motion for $\tilde{Q}_s$ with $s \neq 0$ are independent of degrees 
of freedom in a bath and they are given by
\begin{eqnarray}
M \ddot{\tilde{Q}}_u &=& -M \tilde{\Omega}_u^2 \tilde{Q}_u+ \sqrt{N_S} \:f(t)
\hspace{1cm}\mbox{for $s\neq 0$ and $s = u \neq 0$}, \\
M \ddot{\tilde{Q}}_s &=& -M \tilde{\Omega}_s^2 \tilde{Q}_s
\hspace{3cm}\mbox{for $s\neq 0$ and $s\neq u \neq 0$}.
\end{eqnarray}

The response to applied force with $u$ $(\neq 0)$ is given by
\begin{eqnarray}
R(t) &=& \left( \frac{N_S \:g }{M} \right) 
\int_0^t \frac{\sin \tilde{\Omega}_u(t-t') f(t')}{\tilde{\Omega}_u} \:dt',
\label{eq:G2}
\end{eqnarray}
which becomes for sinusoidal force [Eq. (\ref{eq:E1})],
\begin{eqnarray}
R(t) &=& \left( \frac{N_S \:g }{M} \right) 
\left( \frac{\tilde{\Omega}_u \sin \omega t- \omega \sin \tilde{\Omega}_u t}
{\tilde{\Omega}_u(\tilde{\Omega}_u^2-\omega^2)} \right).
\label{eq:G3}
\end{eqnarray}
In the limit of $c=0.0$, $R(t)$ is given by Eqs. (\ref{eq:G2}) and (\ref{eq:G3})
with $\tilde{\Omega}_u = \Omega_u$.
The effect of finite coupling is realized by a change
in $\tilde{\Omega}_u$ as given by Eq. (\ref{eq:C11}).
Note that the response to applied force with $u=0$ has been studied
in subsection II D [Eq. (\ref{eq:D12})].

We present model calculations for sinusoidal forces with $\omega=0.5$ and $g=1.0$
in Eq. (\ref{eq:E1}) for $N_S=12$, $N_B=100$, $K=B=M=m=\omega_0=1.0$.
Figure \ref{fig5} shows $R(t)/N_S$ for isolated systems ($c_0=0.0$)
with $u=0.0$, $-2.0$, $-4.0$ and $-6.0$.  
Magnitudes of $R(t)/N_S$ become smaller for larger $\vert u \vert$.
Figure \ref{fig6}(a) and \ref{fig6}(b) show $R(t)/N_S$ for uniform ($u=0.0$) and
staggered forces ($u=-6.0$), respectively,
with couplings of $c_0=0.0$ (dashed curve), 5.0 (chain curve) and 10.0 (solid curve).
Comparing Fig. \ref{fig6}(b) with Fig. \ref{fig6}(a), we notice
that an effect of couplings for staggered forces is less effective than
that for uniform forces.

\subsection{Spatial correlation}
Employing eigenfrequencies for $f(t)=0$ obtained in subsection II C, we may calculate
the spatial correlation between $Q_k$ and $Q_{k+m}$,
\begin{eqnarray}
\Gamma_{m} &\equiv& \sum_{k=1}^{N_S} \left<Q_{k}Q_{k+m} \right>, 
\label{eq:F1}\\
&=& \sum_{s=-N_S/2}^{N_S/2-1} 
\left<\tilde{Q}_{s}^{*} \tilde{Q}_{s} \right> e^{- i \:2 \pi m s/N_S},
\label{eq:F2}
\end{eqnarray}
with $\langle \tilde{Q}_{s}^{*} \tilde{Q}_{s} \rangle$ evaluated by
\cite{Grabert84}
\begin{eqnarray}
\left<\tilde{Q}_{s}^{*} \tilde{Q}_{s} \right> 
&=& 
-\left( \frac{1}{\beta M \tilde{\Omega}_s} \right)
\frac{\partial \ln Z_S}{\partial \tilde{\Omega}_s}, 
\label{eq:F3}
%
\end{eqnarray}
where the bracket $\langle \rangle$ denotes the average over $H$ and
$Z_S$ is given by Eq. (\ref{eq:C20}).
$\Gamma_m$ with $m=0$ expresses a (summed) variance of $Q_k$:
$\Gamma_0=\sum_{k=1}^{N_S} \langle Q_k^2\rangle$.
After some manipulations with the use of the diagonalized Hamiltonian
given by Eq. (\ref{eq:C15}), we obtain
\begin{eqnarray}
\left< \tilde{Q}_s^* \tilde{Q}_s \right>
&=& \frac {\hbar}{2 M \tilde{\Omega}_s} 
\coth \left( \frac{\beta \hbar \tilde{\Omega}_s }{2} \right)
\hspace{3cm}\mbox{for $s \neq 0$}, 
\label{eq:F5}\\
&=& \frac {\hbar}{2 M \tilde{\Omega}_0} \sum_{i=1}^2 
\coth \left( \frac{\beta \hbar \phi_i }{2} \right)
\left( \frac{\partial \phi_i}{\partial \tilde{\Omega}_0} \right) 
\hspace{1cm}\mbox{for $s = 0$},
\label{eq:F6}
\end{eqnarray}
with 
\begin{eqnarray}
\frac{\partial \phi_1}{\partial \tilde{\Omega}_0} 
&=& \frac{\tilde{\Omega}_0}{2 \phi_1}
\left[ 1 + \frac{\tilde{\Omega}_0^2-\omega_0^2}
{\sqrt{(\tilde{\Omega}_0^2-\omega_0^2)^2+ 4 N_S N_B c^2/M m}} \right], \\
\frac{\partial \phi_2}{\partial \tilde{\Omega}_0} 
&=& \frac{\tilde{\Omega}_0}{2 \phi_2}
\left[ 1 - \frac{\tilde{\Omega}_0^2-\omega_0^2}
{\sqrt{(\tilde{\Omega}_0^2-\omega_0^2)^2+ 4 N_S N_B c^2/M m}} \right],
\end{eqnarray}
where $\phi_1$ and $\phi_2$ are given by Eqs. (\ref{eq:C16}) and (\ref{eq:C17}).
Substituting Eqs. (\ref{eq:F5}) and (\ref{eq:F6}) into Eq. (\ref{eq:F4}), we obtain $\Gamma_m$,
\begin{eqnarray}
\Gamma_m &=& \sum_{s=-N_S/2}^{N_S/2-1} \frac{\hbar}{2 M \tilde{\Omega}_s}
\coth \left( \frac{\beta \hbar \tilde{\Omega}_s}{2} \right)\:e^{-i 2 \pi m s/N_s} \nonumber \\
&+& \frac {\hbar}{2 M \tilde{\Omega}_0} \left[ \sum_{i=1}^2 
\coth \left( \frac{\beta \hbar \phi_i }{2} \right)
\left( \frac{\partial \phi_i}{\partial \tilde{\Omega}_0} \right) 
- \coth \left( \frac{\beta \hbar \tilde{\Omega}_0}{2} \right)\right].
\end{eqnarray}
For $T=0$ and $T \rightarrow \infty$, $\Gamma_m$ becomes
\begin{eqnarray}
\Gamma_m &=& \sum_{s=-N_S/2}^{N_S/2-1} \left( \frac{\hbar}{2 M \tilde{\Omega}_s} \right)
\:e^{-i 2 \pi m s/N_s} 
+ \frac {\hbar}{2 M \tilde{\Omega}_0} \left[ \sum_{i=1}^2 
\left( \frac{\partial \phi_i}{\partial \tilde{\Omega}_0} \right) 
- 1\right]
\hspace{0.5cm}\mbox{for $T=0$}, 
\label{eq:F7} \\
&=& \sum_{s=-N_S/2}^{N_S/2-1} \left( \frac{k_B T}{M \tilde{\Omega}_s^2} \right)
\:e^{-i 2 \pi m s/N_s} 
+  \frac {k_B T}{ M \tilde{\Omega}_0} \left[ \sum_{i=1}^2 
\frac{\partial \ln \phi_i}{\partial \tilde{\Omega}_0}
- \frac{1}{\tilde{\Omega}_0}\right]
\hspace{0.5cm}\mbox{for $T \rightarrow \infty$}.
\label{eq:F8}
\end{eqnarray}
In the case of uncoupled, isolated system with $K=0.0$ and $c=0.0$, 
$\Gamma_m$ is given by
\begin{eqnarray}
\Gamma_{m} &=& \delta_{m 0} \left( \frac{N_S \hbar}{2 M \tilde{\Omega}_0} \right)
\coth \left(\frac{\beta \hbar \tilde{\Omega}_0}{2} \right)
\hspace{0.5cm}\mbox{for $K=0.0$ and $c_0=0.0$},
\end{eqnarray}
which is proportional to $N_S$ and which vanishes for $m \geq 1$. 
It is, however, not the case for $K \neq 0.0$ or $c \neq 0.0$. 
Indeed in the case of $K \neq 0.0$, $\Gamma_m$ is finite for $m \geq 1$ 
because of direct particle-particle couplings of $K$ and indirect couplings 
of $-c_{k \ell}c_{\ell n}/m \tilde{\omega}_n^2$ in the second term
of Eq. (\ref{eq:A9}). Even when $K=0.0$, $\Gamma_m$ with $c \neq 0.0$
remains finite with a small negative value. 

\begin{figure}
\begin{center}
\includegraphics[keepaspectratio=true,width=120mm]{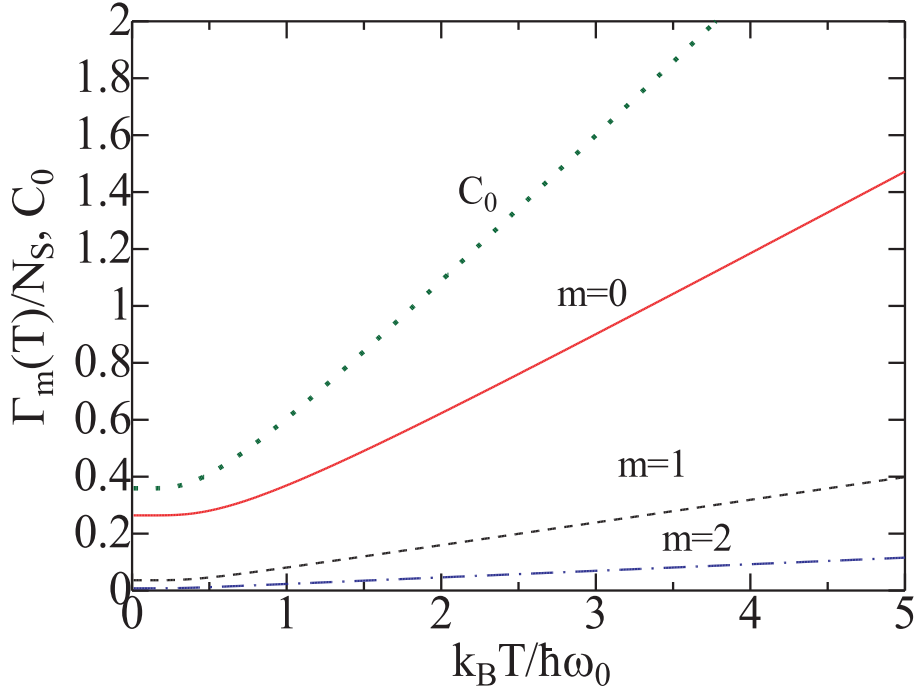}
\end{center}
\caption{
(Color online) 
The temperature dependence of $\Gamma_{m}(T)/N_S$ for $m=0$ (solid curve), 
$m=1$ (dashed curve), $m=2$ (chain curve) and 
$C_0$ ($=\langle \tilde{Q}_0^* \tilde{Q}_0 \rangle$) (dotted curve)   
for a HO system ($N_S=10$, $N_B=100$, $K=D=M=m=\omega_0=1.0$ 
and $c_0=10.0$)
}
\label{fig7}
\end{figure}

\begin{figure}
\begin{center}
\includegraphics[keepaspectratio=true,width=120mm]{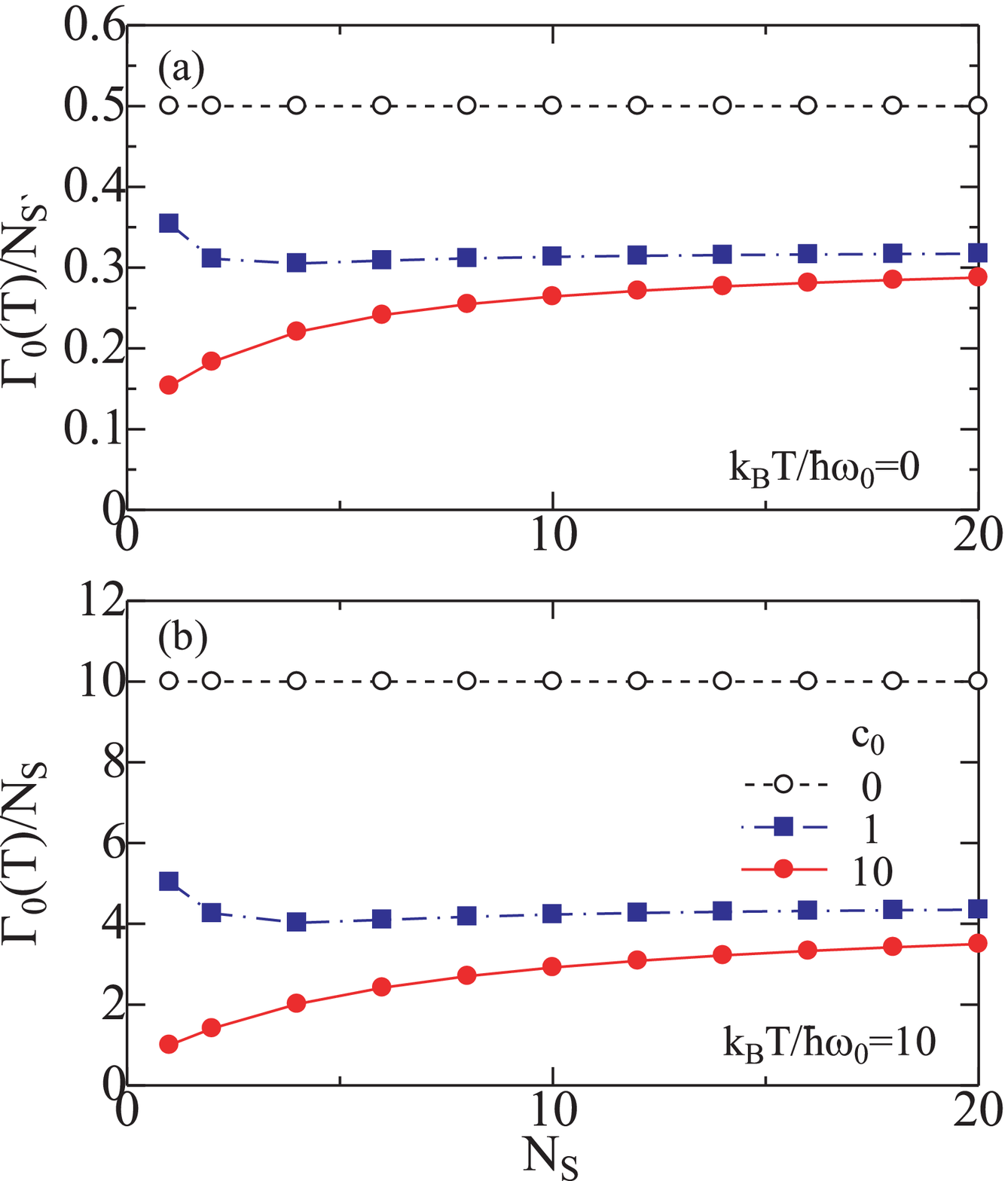}
\end{center}
\caption{
(Color online)
The $N_S$ dependence of $\Gamma_{0}(T)/N_S$ at (a) $k_B T/\hbar \omega_0=0.0$ 
and (b) $k_B T/\hbar \omega_0=10.0$ for $c_0=0.0$ (dashed curve), $1.0$ (chain curve) 
and $10.0$ (solid curve) of a HO system ($N_S=10$, $N_B=100$, $K=D=M=m=\omega_0=1.0$).
}
\label{fig8}
\end{figure}

Figure \ref{fig7} shows the temperature dependence of $\Gamma_m(T)/N_S$ 
for $m=0$ (solid curve), 1 (dashed curve) and 2 (chain curve) 
of HO systems with $N_S=10$, $N_B=100$, $K=D=M=m=\omega_0=1.0$ 
and $c_0=10.0$. $\Gamma_m(T)$ is finite at $T=0$, and at $T \rightarrow \infty$
it is proportional to temperature, as Eqs. (\ref{eq:F7}) and (\ref{eq:F8}) show.
Magnitude of $\Gamma_m$ is smaller for a larger $m$.
The dotted curve expresses 
$C_0$ ($=\langle \tilde{Q}_0^* \tilde{Q}_0 \rangle$) which is larger
than $\Gamma_0$ because $\tilde{\Omega}_0 < \tilde{\Omega}_s$ with
$s \neq 0$.

$N_S$ dependences of $\Gamma_0(T)/N_S$ at $k_B T/\hbar \omega_0=0.0$ and $10.0$ are shown 
in Fig. \ref{fig8}(a) and \ref{fig8}(b), respectively,
for $c_0=0.0$ (open circles), 1.0 (filled square) and 10.0 (filled circles).
For $c_0=0.0$, $\Gamma_0(0)$ is proportional to $N_S$ as expected.
However, when the system-bath coupling is introduced, $\Gamma_0$ is not proportional
to $N_S$ as shown in Fig. \ref{fig8}.
This is realized not only at zero temperature but also at high temperature.

Even when external forces are applied, the spatial correlation
is not modified, which is the characteristics of the open system
with the linear system-bath coupling. 
In the open system with the nonlinear system-bath coupling, the spatial correlation
is modified by an applied force \cite{Hasegawa11d}.

\section{Conclusion}
Responses of open small quantum systems described by the $(N_S+N_B)$ model
\cite{Hasegawa11a,Hasegawa11b}
have been studied.  By using double canonical transformations mentioned in
subsections II B and II C,
we obtain the diagonalized Hamiltonian, from which the response to
applied forces is obtained with the use of Husimi's method 
for a driven quantum HO \cite{Husimi53}. 
The response to a uniform force given by Eq. (\ref{eq:A2b}) 
is generally not proportional to $N_S$ against our implicit expectation.  
This nonlinear response is consistent with the system specific heat
in open small quantum systems previously discussed in Ref. \cite{Hasegawa11b}, 
and it is realized also in spatial correlation $\Gamma_m$ not only
at low temperatures but also at high temperatures.
These facts show an importance of taking account of finite $N_S$
in discussing open quantum and classical systems.
It would be interesting to examine the obtained non-linearly by experiments 
for open small systems.

\begin{acknowledgments}
This work is partly supported by
a Grant-in-Aid for Scientific Research from 
Ministry of Education, Culture, Sports, Science and Technology of Japan.  
\end{acknowledgments}


\end{document}